\documentclass[twocolumn,showpacs,preprintnumbers,amsmath,amssymb,pra,unsortedaddress,superscriptaddress,longbibliography]{revtex4-1}
\usepackage{graphicx}
\usepackage{dcolumn}
\usepackage{bm}
\usepackage{verbatim} 
\usepackage{braket}
\usepackage{pbox}
\usepackage{multirow}
\usepackage{array}
\usepackage[english]{babel}
\usepackage{letltxmacro}

\LetLtxMacro{\ORIGselectlanguage}{\selectlanguage}
\makeatletter
\DeclareRobustCommand{\selectlanguage}[1]{%
  \@ifundefined{alias@\string#1}
    {\ORIGselectlanguage{#1}}
    {\begingroup\edef\x{\endgroup
       \noexpand\ORIGselectlanguage{\@nameuse{alias@#1}}}\x}%
}
\newcommand{\definelanguagealias}[2]{%
  \@namedef{alias@#1}{#2}%
}
\makeatother

\definelanguagealias{en}{english}

\newcommand{\pvec}[1]{\vec{#1}\mkern2mu\vphantom{#1}}

\begin{document}


\title{Simulations of the angular dependence of the dipole-dipole interaction among Rydberg atoms}

\author{Jacob L. Bigelow}
\author{Jacob T. Paul}
\author{Matan Peleg}
\author{Veronica L. Sanford}
\author{Thomas J. Carroll}
\affiliation{Department of Physics and Astronomy, Ursinus College, Collegeville, PA 19426.}

\author{Michael W. Noel}%
\affiliation{Department of Physics, Bryn Mawr College, Bryn Mawr, PA 19010.}

\date{\today}

\begin{abstract}
The dipole-dipole interaction between two Rydberg atoms depends on the relative orientation of the atoms and on the change in the magnetic quantum number. We simulate the effect of this anisotropy on the energy transport in an amorphous many atom system subject to a homogeneous applied electric field. We consider two experimentally feasible geometries and find that the effects should be measurable in current generation imaging experiments. In both geometries atoms of $p$ character are localized to a small region of space which is immersed in a larger region that is filled with atoms of $s$ character. Energy transfer due to the dipole-dipole interaction can lead to a spread of $p$ character into the region initially occupied by $s$ atoms. Over long timescales the energy transport is confined to the volume near the border of the $p$ region which suggests Anderson localization. We calculate a correlation length of 6.3~$\mu$m for one particular geometry. 
\end{abstract}

\maketitle

\section{Introduction}
The energy exchange among Rydberg atoms has garnered great interest both as a probe of fundamental quantum dynamics and as a potential way to model other physical systems. The strong dipolar interactions among Rydberg atoms can block all but one excitation in a group of atoms, leading to a collective two-level system~\cite{lukin_dipole_2001,gaetan_observation_2009,urban_observation_2009}. These ``super-atoms'', which could be used as qubits, have recently been observed and characterized~\cite{zeiher_microscopic_2015,weber_mesoscopic_2015}. The control of Rydberg atoms has also made progress at the single atom level. Coherent coupling between a pair of Rydberg atoms was directly measured revealing both the $1/r^3$ dependence of the dipole-dipole interaction and the $1/r^6$ dependence of the van der Waals interactions~\cite{ravets_coherent_2014,beguin_direct_2013}. Systems of Rydberg atoms have been proposed as quantum simulators of frustrated magnets and ``quantum spin ices''~\cite{glaetzle_quantum_2014,glaetzle_designing_2015}. Coherent excitation hopping has been observed in a chain of three Rydberg atoms, pointing toward the use of Rydberg systems as quantum simulators of spin dynamics~\cite{barredo_coherent_2015}. The energy transport in such systems has been studied in simulation, for example by investigating the effect of randomness on energy transport in a lattice~\cite{robicheaux_effect_2014} or exploring the importance of dissipation and correlations~\cite{schempp_correlated_2015}. Understanding the dynamics of energy transport in systems of cold Rydberg atoms will be important when using these systems for quantum information tasks.


The energy exchange between a pair of Rydberg atoms is mediated by the dipole-dipole interaction,
\begin{equation}\label{eq:angdep}
V(\mathbf{r}) = \frac{\boldsymbol{\mu_1}\cdot\boldsymbol{\mu_2}-3(\boldsymbol{\mu_1}\cdot\mathbf{\hat{R}})(\boldsymbol{\mu_2}\cdot\mathbf{\hat{R}})}{R^3},
\end{equation}
where $\mathbf{R}$ is the separation vector between the atoms and $\boldsymbol{\mu_i}$ are the electric dipole matrix elements connecting the initial and final states for each atom. The external electric field fixes the quantization axis. The dependence of the interaction on $\theta$, the zenith angle between the quantization axis and $\mathbf{\hat{R}}$, arises from the second term in the numerator of Eq.~(\ref{eq:angdep})~\cite{carroll_angular_2004}. If the applied electric field is homogeneous, the three possibilities for the $\theta$ dependence are, for each possible combination of $\Delta m_{j1}$ and $\Delta m_{j2}$:
\begin{multline}
\Delta m_{j1} + \Delta m_{j2} = 0:\\
 f_0(\theta) =  (1-3\cos^2\theta)/2,\label{eq:f0}
\end{multline}
\begin{multline}
\Delta m_{j1} + \Delta m_{j2} = \pm 1:\\
f_1(\theta) = (3\sin\theta\cos\theta)/\sqrt{2},\label{eq:f1}
\end{multline}
\begin{multline}
\Delta m_{j1} + \Delta m_{j2} = \pm 2:\\
f_2(\theta) = (3\sin^2\theta)/2.\label{eq:f2}
\end{multline}

Experimental evidence for the angular dependence of the dipole-dipole interaction was presented in~\cite{carroll_angular_2004} by confining Rydberg atoms to a one dimensional sample. Recently, Ravets \textit{et al.} have directly measured the anisotropy by restricting the dipole-dipole interaction to one channel and considering only two atoms; they find good agreement with the predicted angular dependence~\cite{ravets_measurement_2015}. In a similar system of magnetic spins, the angular dependence of the  dipole-dipole interaction was observed to create an anisotropic deformation in the expansion of a Bose-Einstein condensate of chromium \cite{Stuhler_observation_2005} and dysprosium~\cite {lu_strongly_2011}.  An Erbium condensate exhibited a d-wave collapse when the scattering length was tuned below a critical value using a Feschbach resonance~\cite{aikawa_bose-einstein_2012}. 

A number of recent experiments have studied Rydberg systems by imaging the spatial distribution of Rydberg atoms as a function of time. If the imaging technique is state selective, such experiments can measure the spatial and temporal dependence of the energy transport. Imaging experiments have been used to study correlations due to the dipole blockade~\cite{schwarzkopf_imaging_2011,schwarzkopf_spatial_2013}. The evolution of a cloud of Rydberg atoms toward an ultracold plasma has been imaged by scattering light from core ions~\cite{mcquillen_imaging_2013}. The diffusion of energy throughout an ultracold gas of Rydberg atoms has been observed, with sufficient spatial and temporal resolution to extract a diffusion rate~\cite{gunter_observing_2013}. The exchange of energy between two spatially separated groups of atoms has been imaged~\cite{fahey_imaging_2015}.

Motivated by recent imaging results, we simulate the time evolution of the energy exchange among Rydberg atoms by numerically solving the Schr\"{o}dinger equation. Our model includes the full angular dependence of the dipole-dipole interaction in the many atom regime. With relatively simple and experimentally accessible geometries, we find that it should be possible to observe the effect of the anisotropy on the energy transport.

\section{Model}

\begin{figure}
	\includegraphics{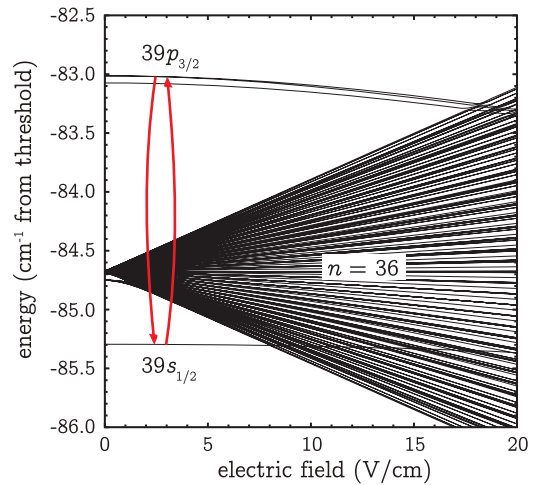}
	\caption{\label{fig:starkmap} (color online) Stark map showing an example of the dipole-dipole interaction $s + p \rightarrow p + s$. This energy exchange is ``always resonant'' provided that the applied static electric field is spatially homogeneous.}
\end{figure}

We consider an amorphous sample of Rydberg atoms in the presence of a static electric field. For concreteness, we focus our attention on the energy exchange
\begin{equation}\label{eq:ddsp}
p_{3/2,|m_j|=3/2} + s_{1/2} \rightarrow s_{1/2} + p_{3/2,|m_j|=3/2},
\end{equation}
which is often called the always resonant channel as the interaction can proceed regardless of the value of the external electric field, though a field inhomogeneity can shift this interaction away from resonance~\cite{fahey_imaging_2015}. An example is shown in the Stark map of Fig.~\ref{fig:starkmap} for Rb $39s_{1/2}$ and $39p_{3/2}$ states. This interaction is advantageous because it has large dipole moments ($\sim$800~$ea_0$) and does not require precise tuning of the electric field.

Since the $s$ atoms have $|m_j|=1/2$ then $\Delta m_j$ cannot be zero for either atom. Thus the sum $\Delta m_{j1} + \Delta m_{j2} = 0$ or $\pm 2$ but the $\pm 1$ case is eliminated. By choosing the direction of an applied static electric field we fix the quantization axis and hence the relative orientation angle of the Rydberg atoms. Because of the anisotropy of Eq.~(\ref{eq:f0}) and Eq.~(\ref{eq:f2}), the energy transport will be more efficient for particular relative orientation angles. The interaction between the atoms will be strongest when $\theta = \pi/2$.

Experimentally, one could selectively excite the $ p_{3/2,|m_j|=3/2}$ state by applying a small electric field, which splits the two $|m_j|$ states.  One caution is that at low applied fields the $p_{3/2},|m_j|=1/2$ state is close in energy to the $|m_j|=3/2$ state. In this case the $s_{1/2}$ state could couple off-resonantly by the interaction
\begin{equation}
p_{3/2,|m_j|=3/2} + s_{1/2} \rightarrow s_{1/2} + p_{3/2,|m_j|=1/2}.
\end{equation}
This would introduce the possibility of $\Delta m_j = \pm 1$ interactions, which would diminish the visibility of the anisotropic energy transport as there would no longer be a preferred angle.

In order to maximize the number of $s$ and $p$ atoms that we can model, we do not include the ground state or interaction with the excitation lasers. Energy exchange during the laser excitation has been shown to play a role in the observed state mixing in some experiments~\cite{younge_mesoscopic_2009}. In our model, we consider relatively short excitation pulses that provide a well-defined starting time for the interaction and minimal opportunity for dipole-dipole energy exchange during excitation.

The Hamiltonian for our system in a homogeneous electric field is
\begin{multline}\label{eq:ham}
\hat{H}_{ij}=\sum_{i\ne j} \frac{\mu^2}{R_{ij}^3}
\biggl[f_0(\theta)\bigl(\hat{\sigma}^i_{p+s+}\hat{\sigma}^j_{s+p+} + 
\hat{\sigma}^i_{p-s-}\hat{\sigma}^j_{s-p-}\bigr) +\\
f_2(\theta)\bigl(\hat{\sigma}^i_{p+s+}\hat{\sigma}^j_{s-p-} + 
\hat{\sigma}^i_{p-s-}\hat{\sigma}^j_{s+p+} \bigr)+ \mathrm{H.c.}\biggr],
\end{multline}
where $\mu$ is the dipole moment connecting the $s$ and $p$ states, $R_{ij}$ is the distance between the $i^{th}$ and $j^{th}$ atoms, and $\hat{\sigma}^i_{x\pm y\pm}$ is an operator that changes the state of the $i^{th}$ atom from $\ket{x\pm}$ to $\ket{y\pm}$. The plus and minus signs denote the positive and negative values of $m_j$ for each state; for example, $\hat{\sigma}_{p+s+}$ takes an atom from $\ket{p_{3/2},m_j=+3/2}$ to $\ket{s_{1/2}, m_j=+1/2}$. 

The Hamiltonian matrix generated by Eq.~(\ref{eq:ham}) will be quite large even for relatively small numbers of atoms because we must include all possible $m_j$ states for each atom. The size of the matrix is
\begin{equation}
N = {n \choose n_s}2^n,
\end{equation}
where $n$ is the total number of atoms and $n_s$ is the number of $s$ atoms. We consider our initial state to be a superposition of all possible $m_j$ states for our initially excited population of $s$ and $p$ atoms. Since we are modeling an amorphous sample and wish to average over both space and time, we simulate thousands of instances. With available local and NSF XSEDE~\cite{towns_xsede:_2014} supercomputing resources, we can achieve this for matrices $N \lesssim 10^5$. Given this practical limit, if we simulate 3 $p$ atoms we can include at most 6 additional $s$ atoms. For one or two $p$ atoms we can include up to 11 or 8 $s$ atoms, respectively. In this work, we include cases the following cases: one $p$ atom with 6, 7, 8, and 9 $s$ atoms and two $p$ atoms with 5, 6, 7, and 8 $s$ atoms.

\section{Results and Discussion}

We model two geometries, both of which localize a small number of $p$ atoms in a larger group of $s$ atoms. We consider a sphere of $p$ atoms localized at the end of a long and narrow cylinder of $s$ atoms. This restricts the energy transport to a roughly one dimensional geometry. We also consider a thin disc of $s$ atoms with a small region of $p$ atoms in the center, which allows us to image energy transport in two dimensions.

In each case we simulate the time evolution in steps of $<0.1$~$\mu$s. At each time step, we store the probability for each atom to be in either the $s$ or the $p$ state. We average the results of multiple runs together by binning the probabilities by the atom's position. This generates a map of the energy transport as a function of position, as shown in Fig.~\ref{fig:cyl} and Fig.~\ref{fig:disc}. 

\subsection{One dimensional case: cylindrical geometry}

We randomly place $p$ atoms in a spherical volume of diameter 10~$\mu$m. The spherical volume is at one end of a cylinder of the same diameter and length 60~$\mu$m, in which we randomly place $s$ atoms. Such a configuration could be achieved with a multi-step excitation such as the one described in~\cite{fahey_excitation_2011}. The overlap of two perpendicular laser beams used to excite $p$ atoms would define a localized volume represented by our spherical volume. Another laser beam, collinear with one of the $p$-excitation beams and exciting $s$ atoms, would define a roughly cylindrical volume.

Since $\Delta m_j \ne 0$, the angular dependence of Eq.~(\ref{eq:f1}) is not possible and the only $\theta$ dependence is from Eq.~(\ref{eq:f0}) or Eq.~(\ref{eq:f2}). If the electric field is perpendicular to our cylinder, then $\theta$ for a given pair of atoms is likely to be near $\pi/2$. At this angle the magnitude of Eq.~(\ref{eq:f0}) is $f_0(\pi/2)=1/2$ and the magnitude of Eq.~(\ref{eq:f2}) is at a maximum value $f_2(\pi/2)=3/2$. 

If the electric field is parallel to our cylinder, then $\theta$ for a given pair of atoms is likely to be near zero. At this angle the magnitude of Eq.~(\ref{eq:f0}) is $f_0(\pi/2)=-1$ while the magnitude of Eq.~(\ref{eq:f2}) is at a minimum value $f_2(\pi/2)=0$. The contribution of the angular dependence to the overall coupling between the atoms is on average larger for angles near $\pi/2$. Thus, we expect that for a perpendicular field we should have more efficient energy transport along the cylinder.

\begin{figure}
	\includegraphics{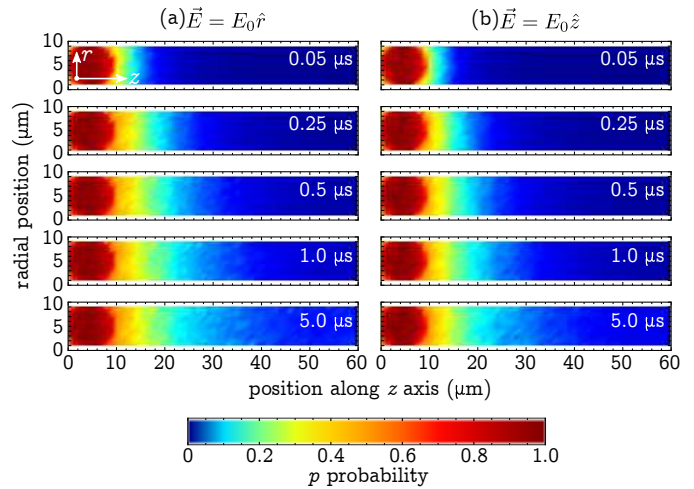}
	\caption{\label{fig:cyl} (color online)  Expansion of $p$ character from the initial spherical volume at left into the cylindrical volume initially populated with only $s$ atoms. The probability of detecting a $p$ atom at a position is normalized to the number of atoms excited at that position over all simulation runs. The color scale is given by the legend, with redder colors indicating more $p$ character and bluer colors indicating less $p$ character. Horizontally adjacent images are generated at the same time step for direct comparison. (a) The expansion when an electric field $E_0\hat{r}$ (perpendicular to the beam) is applied. (b) The expansion when an electric field $E_0\hat{z}$ (along the beam) is applied. The expansion in the axial case is evidently slower than the expansion in the perpendicular case due to the angular dependence of the energy exchange. Note that the color scale has been chosen to enhance contrast at low probability.}
\end{figure}

We run our simulation both for the case of an electric field parallel to the cylindrical excitation volume, $\vec{E}=E_0\hat{z}$, and an electric field perpendicular to the cylindrical excitation volume, $\vec{E}=E_0\hat{r}$. The results are shown in Fig.~\ref{fig:cyl}, where all different simulation cases are averaged. Each column in Fig.~\ref{fig:cyl} shows the diffusion of $p$ character along the cylinder at four different times. The diffusion is clearly faster in the case of a perpendicular field (Fig.~\ref{fig:cyl}(a)) as compared to the case of a parallel field (Fig.~\ref{fig:cyl}(b)). 

It is interesting to note that there is not much difference in the total energy exchange between these two cases. At positions near the right end of the cylinder, there is more $p$ character in Fig.~\ref{fig:cyl}(a). For positions near the left end of the cylinder, there is more $p$ character in Fig.~\ref{fig:cyl}(b). This is because pairs of atoms comprised of a $p$ atom in the initially populated sphere and a nearby $s$ atom will not have their relative orientation angle constrained by the cylindrical geometry. In both cases of field direction, these pairs of atoms should exchange energy efficiently. However, for the case of Fig.~\ref{fig:cyl}(a), initial $s$ atoms near the left end of the cylinder that have mixed with significant $p$ character can more efficiently transport their energy farther along the cylinder.

One of the striking features of Fig.~\ref{fig:cyl}(a) and (b) is that the energy transport does not proceed very far down the cylinder. The color scale is chosen to enhance the contrast at the low probability end of the scale; considering this there is evidently little change between $t=0.5~\mu$s and $t=5.0~\mu$s. 

\begin{figure}
	\includegraphics{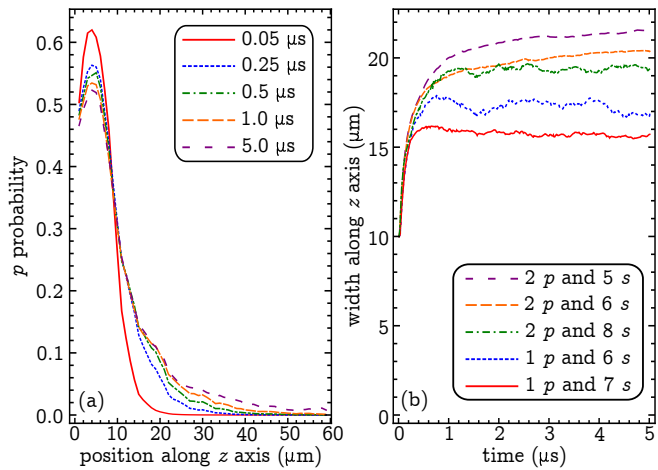}
	\caption{\label{fig:aloc} (color online)  (a) Integrated probability of detecting a $p$ atom as a function of position along the length of the cylinder for the case which includes 2 $p$ atoms and 8 $s$ atoms. Most of the expansion of $p$ probability happens in the first microsecond, with little subsequent change. (b) Width as a function of time for the distribution shown in (a) and other selected cases. The width is measured at $0.2~\times$ maximum height so as to be sensitive to the small changes in the $p$ probability at large $z$. The rapid change at early times plateaus sharply suggesting localization. The electric field in both (a) and (b) is in the $\hat{r}$ direction, as in Fig.~\ref{fig:cyl}(a).}
\end{figure}

We examine this more quantitatively in Fig.~\ref{fig:aloc}, where the electric field is in the transverse ($\hat{r}$) direction for all cases. For Fig.~\ref{fig:aloc}(a) we integrate the $p$ probability across the cylinder at each $z$ and this integrated $p$ probability is shown for five different times for the simulated case of 2~$p$ atoms and 8~$s$ atoms. Most of the increase in the width of the distribution occurs for $t<1$~$\mu$s. This is clearly displayed in Fig.~\ref{fig:aloc}(b) for a variety of selected cases. In all cases the width rapidly increases at early times and then sharply plateaus. While we have explored only a small region of parameter space for low numbers of atoms, a trend is visible in these results. The cases in which the ratio of $s$ atoms to $p$ atoms is largest display the flattest plateaus, while those cases with a smaller ratio still have a slight upward trend at later times.

\begin{figure}
	\includegraphics{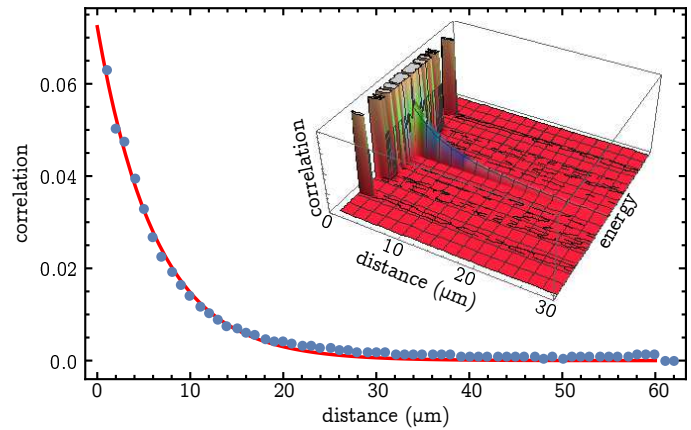}
	\caption{\label{fig:loclen} (color online) The intensity-intensity correlation of Eq.~(\ref{eq:corr}) for the narrow band of eigenenergies in the middle of the inset, which show some extended correlation. The data here are from the 1~$p$ and 7~$s$ atom case for the cylindrical geometry discussed in the text. The solid red line is a fit to an exponential decay which yields a localization length of $\sim$6.3~$\mu$m. }
\end{figure}

The results of Fig.~\ref{fig:cyl} and Fig.~\ref{fig:aloc} suggest Anderson localization~\cite{anderson_absence_1958} of the $p$ character. In order to characterize the localization observed here, we numerically calculate the intensity-intensity correlation function
\begin{equation}
\langle |\psi_E(\vec{r})|^2 |\psi_E(\pvec{r}')|^2 \rangle\label{eq:corr}
\end{equation}
for each eigenfunction $\psi_E(\vec{r})$ of our system. We bin the results by energy and position from $\sim 10^4$ simulation runs for the 1~$p$ and 7~$s$ case. A portion of the resulting graph is shown in the inset of Fig.~\ref{fig:loclen}, which shows the correlation of Eq.~(\ref{eq:corr}) as a function of the distance $|\vec{r} -\pvec{r}'|$ and the binned eigenenergies. Only a narrow band of energies show any extended correlation; these data are graphed as the blue points in Fig.~\ref{fig:loclen}. The solid red line in Fig.~\ref{fig:loclen} is a fit to the function
\begin{equation}
A~ \exp\Bigl(-\frac{|\vec{r}-\pvec{r}'|}{\xi(E)}\Bigr),
\end{equation}
where $A$ is a constant and $\xi(E)$ is the localization length. The fit in Fig.~\ref{fig:loclen} yields a localization length of $\xi \sim 6.3$~$\mu$m.

Robicheaux and Gill~\cite{robicheaux_effect_2014} have investigated the effects of randomness on energy transport in one, two, and three-dimensional lattices of $s$ atoms with one $p$ atom. They randomly perturbed the positions of the atoms or the filling of the lattice and then systematically studied the effect of randomness by examining the energy transport for a range of perturbation sizes. In the one-dimensional case, they find that the $p$ excitation is localized for every eigenstate even for weak randomness. The degree of localization was found to increase with the size of the random perturbation. Our amorphous system is a similar geometry to their maximally random case and our results are in general agreement with Robicheaux and Gill, though our system includes some additional randomness due to the angular dependence and the variable relative orientations of the atoms.

\subsection{Two dimensional case: disc geometry}

We randomly place $s$ atoms in a disc of diameter 60~$\mu$m and thickness 5~$\mu$m and $p$ atoms only in the central region of the disc with diameter 10~$\mu$m. We examine the energy transport when the applied electric field points perpendicular to the disc in the $z$ direction and when it points parallel to the plane of the disc in either the $x$ or $y$ direction. 

\begin{figure}
	\includegraphics{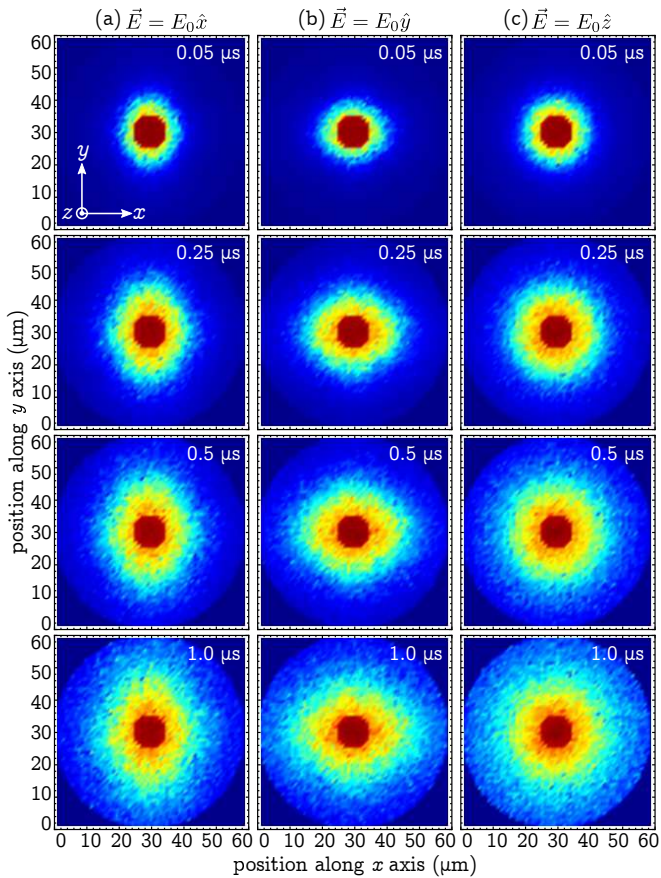}
	\caption{\label{fig:disc} (color online) Expansion of $p$ character from the initial central volume into the planar disc initially populated with only $s$ atoms. The color scale is given by the legend in Fig.~\ref{fig:cyl}, with redder colors indicating more $p$ character and bluer colors indicating less $p$ character. Horizontally adjacent images are generated at the same time step for direct comparison. The orientation of the $x$, $y$, and $z$ axes is given by the inset at the upper left. The disc lies in the $xy$ plane. (a) The expansion when an electric field $E_0\hat{x}$ is applied. (b) The expansion when an electric field $E_0\hat{y}$ is applied. (c) The expansion when an electric field $E_0\hat{z}$ is applied. In (a) the diffusion of $p$ character is faster in the $y$ direction, while in (b) it is faster in the $x$ direction. This is consistent with the angular dependence of $f_0(\theta)$ and $f_2(\theta)$ as described in the main text.}
\end{figure}

The results of the simulation are shown in Fig.~\ref{fig:disc}, where all simulation cases are averaged. Each column shows the diffusion of $p$ character away from the central region for the case of a uniform electric field pointing in the $x$ (Fig.~\ref{fig:disc}(a)), $y$ (Fig.~\ref{fig:disc}(b)), or $z$ direction (Fig.~\ref{fig:disc}(c)). In Fig.~\ref{fig:disc}(a) and (b) the diffusion of energy is asymmetric. This is because  when the field points in the $x$ or $y$ direction, different pairs of atoms will have different relative orientation angles. When the field points in the $x$ ($y$) direction, the maximum rate of diffusion will be in the $y$ ($x$) direction. Similar to the previous discussion of energy transport along the cylinder, this is because the angular dependence will be a maximum when the relative orientation angle is $\pi/2$. When the field points in the $z$ direction, the relative orientation angle $\theta$ between any pair of atoms is always $\pi/2$. The angular dependence will not distinguish any direction of energy diffusion in the plane of the disc. This isotropic diffusion is evident in Fig.~\ref{fig:disc}(c).

\section{Conclusion}

Previous investigations have confirmed the anisotropic nature of the dipole-dipole interaction between pairs of atoms.  We have investigated the consequence of this anisotropic interaction by simulating the dynamics of energy transport in a cold gas of Rydberg atoms. Our model solves the Schr\"{o}dinger equation numerically and includes the full many body wave function. We find that the rate of energy transport depends both on the geometry of the atomic sample and on the angular dependence of the energy exchange as determined by direction of the applied electric field and the various $\Delta m_j$ combinations possible in the interaction. In our pseudo-one-dimensional model, we see the signature of Anderson localization and are able to extract a localization length for some cases. Our simulation results show that current generation imaging experiments should be able to measure the effects of the anisotropy in accessible atomic geometries. This anisotropy  could be used to tune the energy transport in a cold Rydberg gas, making these systems useful as a quantum simulators.

This work was based upon work supported by the National Science Foundation under Grants No. 1205895 and No. 1205897.

This work used the Extreme Science and Engineering Discovery Environment (XSEDE), which is supported by National Science Foundation grant number OCI-1053575.
\bibliography{AngDepSim}

\end{document}